\documentstyle[prl,aps,graphicx,multicol]{revtex}
\begin{document}

\title{Microspectroscopy and Imaging in the THz Range Using Coherent CW Radiation}
\author{S. Mair, B. Gompf,  and M. Dressel\footnote{To whom correspondence should be addressed (dressel@pi1.physik.uni-stuttgart.de)}}
\address{1. Physikalisches Institut, Universit\"{a}t Stuttgart,
Pfaffenwaldring 57, D-70550 Stuttgart, Germany}
\maketitle

\begin{abstract}
A novel THz near-field spectrometer is presented which allows to
perform biological and medical studies with high spectral resolution combined with a spatial resolution
down to $\lambda$/100. In the setup an aperture much smaller than the used wave\-length is placed in the beam
very close to the sample. The sample is probed by the evanescent wave behind the aperture.
The distance is measured extremely
accurate by a confocal microscope. We use monochromatic sources
which provide powerful coherent cw radiation tuneable from 50~GHz
up to 1.5~THz. Transmission and reflection experiments can be
performed which enable us to study solids and molecules in aqueous
solution. Examples for spectroscopic investigations on biological
tissues are presented.
\end{abstract}
\begin{multicols}{2}
\columnseprule 0pt

\section{Introduction}
During recent years there have been many efforts to understand in
detail how cells function. To cope this complex problem, it is
necessary to examine the components of the system, like DNA,
bio-molecules, proteins, cell-membrans. Due to the large entities
and the high density of electronic, vibrational and rotational
states in these bio-molecules, spectroscopic investigations in the
millimeter, submillimeter, and far-infrared spectral range (often
called THz range) seem to be an appropriate method for this
task. However, the dimensions of the samples under study are
rather small, often well below Abbe's diffraction limit, implying
that in the millimeter to far-infrared spectral range it is not
possible to perform measurements  by standard optical methods. In
addition to the scientific point of view, the THz range of
frequency attracts considerable attention because of possible
applications, in particular with the aim of THz imaging. While a
number of groups try to peruse time domain technique utilizing
short laser pulses in order to generate a broad spectrum terahertz
radiation\cite{Hu95,Mittleman99}, we improved the well developed technique in the
frequency domain. Since most biological and
medical investigations will required measurements of molecules in
aqueous solution and water is highly absorbing in the THz range,
we have developed a reflection setup in addition to simple
transmission arrangements.

The frequency range between microwaves and infrared has proven to
be particularly challenging to experimentalists since it falls
right between the range of guided waves (coaxial and microwave
technique) and optical techniques (free space propagation).
Microwave oscillators are in general not tunable over a larger
range of frequency and/or deliver only a very small output power;
hence they basically cannot be employed above 100~GHz. Coming from
optical methods, standard infrared spectroscopy using Fourier
transform interferometer has to cope with two problems by trying
to go to lower frequencies: (i) the output power of the thermal
radiation sources vanishes for $\omega\rightarrow 0$ following
Planck's law of black body radiation; (ii) the spot size to which
the light can be focused for measurements is diffraction limited
to approximately the wavelength. Here we present a new approach to
overcome this limitations by utilizing monochromatic but tunable
backward wave oscillators as powerful radiation sources and a
near-field microscope in order to perform microspectrometry in the
THz range.

\section{Coherent Source  Spectrometer}
\label{sec:spectrometer}
\begin{figure}
\centering\includegraphics[width=87mm]{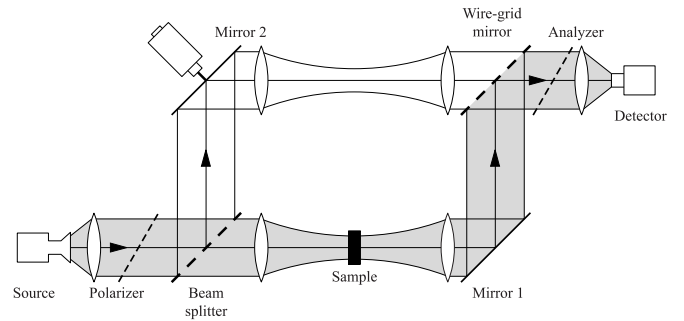}\vspace{3mm}
\caption{\label{fig:MachZehnder}
Mach-Zehnder type interferometer used for quasi-optical transmission measurements in
the THz range of frequency. The coherent radiation generated by
backward wave oscillators is split by wire grids. The length of the
interference arm can be adjusted by moving mirror 2. After the sample is introduced, the
interferometer is readjusted in order to obtain the phase difference.}
\end{figure}
The use of backward wave oscillators as radiation sources for spectroscopic measurement in
the frequency range from 1~cm$^{-1}$\ to 50~cm$^{-1}$\ ($f=30$~GHz to 1.5~THz; $\lambda=0.2$~mm to 10~mm) was
developed in the group of G. Kozlov during the last twenty years \cite{Volkov85,Kozlov98}.
With a cw output power of 1 to 200~mW reliable optical experiments can be performed with a
very short data acquisition time. The radiation can be tuned in frequency with an accuracy
and stability of $10^{-6}$ allowing to map very narrow absorption lines.
The radiation is detected either by a Golay cell or  by a silicon bolometer,
which is cooled down to 1.2~K. The spectrometer has a dynamical range of up to $10^7$. Experiments can be performed
either in reflection or transmission, where a Mach-Zehnder interferometer (Fig.~\ref{fig:MachZehnder}) also allows to probe
the change in phase and thus to evaluate the real and imaginary part of
the electrodynamic response without performing a Kramers-Kronig analysis \cite{DresselGruner}.
Since a quasi-optical arrangement is limited by diffraction and standing-wave problems, samples smaller
than approximately three times the wavelength cannot be investigated.

\section{Near-Field Setup}
\begin{figure}
\centering\includegraphics[width=87mm,clip]{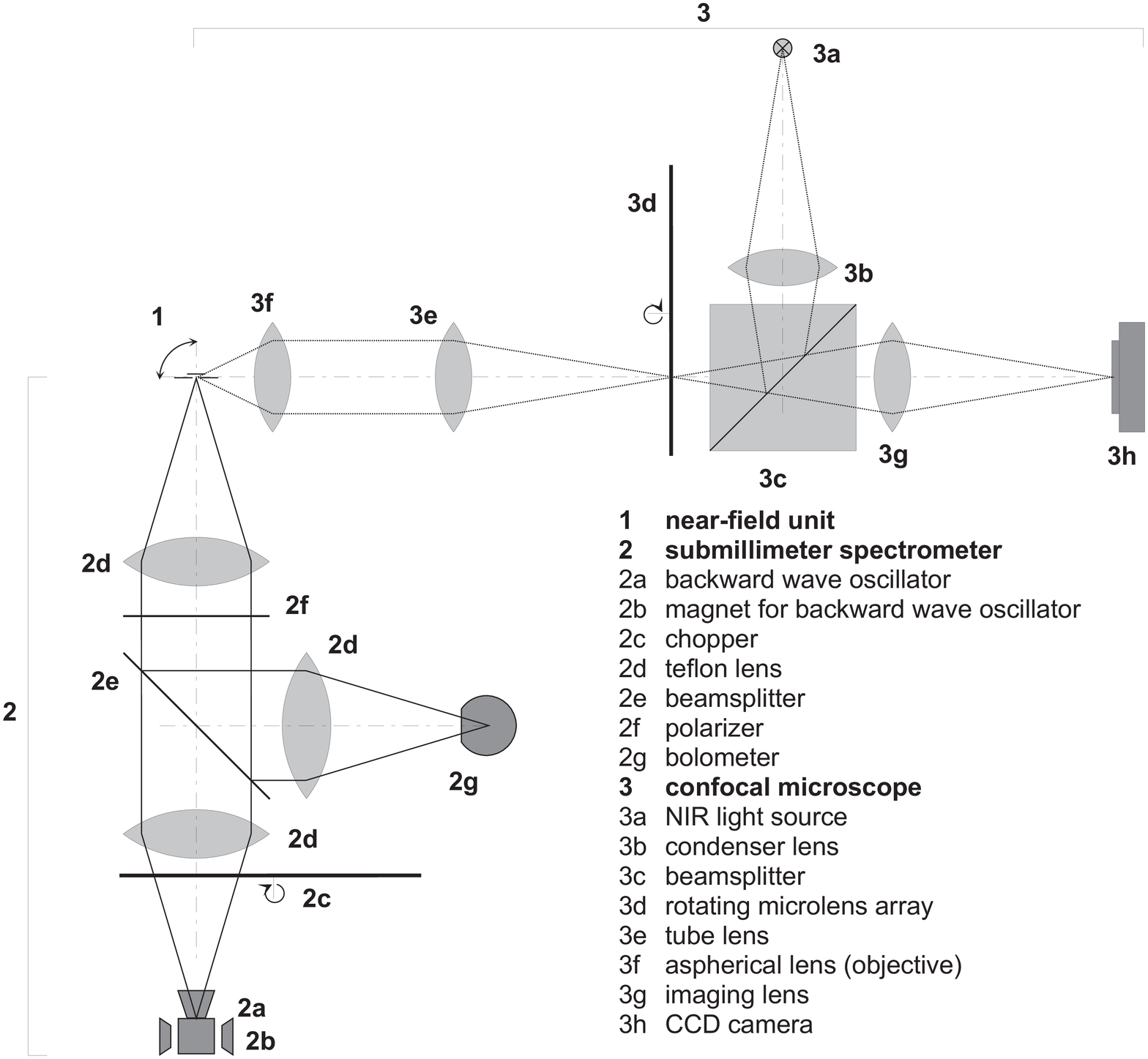}\vspace{3mm}
\caption{\label{fig:scheme} Schematic of the THz near-field
spectrometer. The sample is placed in the near-field unit (1); the
spectroscopic measurements are done by a coherent source
spectrometer (2); the distance control and measurement is done by
a confocal micorscope (3).}
\end{figure}

We have developed a THz near-field microscope which allows us to perform
reflection measurement on  spot sizes down to $\lambda/100$. As displayed in Fig.~\ref{fig:scheme}, the setup consists of a
near-field unit mounted rotatable by 90 degrees (1); one axis is used for a
quasi-optical spectrometer (2), and the second by a confocal microscope (3) for distance control.
The spectroscopic investigations are done with a coherent source
spectrometer similar to the one discribed above in
Sec.~\ref{sec:spectrometer} just arranged for reflection
measurements. The cw beam generated by backward wave oscillators
is mechanically chopped and focussed on the pinhole by teflon
leses. Within a few seconds it can be electronically tuned over
the full range of the oscillator. The reflected beam is guided to
the bolometer by a wire-grid beam splitter.
\begin{figure}
\centering\includegraphics[width=40mm]{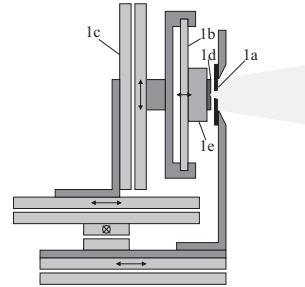}\vspace{3mm}
\caption{\label{fig:nfunit}
Schematic of the near-field unit, consisting of pinhole (1a), piezo bender (1b), piezo motors (1c), sample (1d), and sample holder (1e).}
\end{figure}

The core part of the near-field spectrometer is a small pinhole(1)
which is
drilled in a copper foil (thickness 1 to 10 $\mu$m) with diameter as small as 1~$\mu$m
and which generates the evanescent submillimeter wave the sample is probed with (Fig.~\ref{fig:nfunit}).
Behind the pinhole sits the sample (1d) which can be translated
relative to the pinhole by three linear piezo-motors (1c). These motors are based on
stick and slip motion with a step size of 40 to 400~nm. There is a fourth motor for moving the sample and pinhole together, which is
required for the distance control via the confocal microscope.
The distance between pinhole and sample is modulated up to $\pm$70 $\mu$m
with high frequencies (about 1 kHz) by an
additional piezo-electric bender (1b) in order to using look-in technique for data acquisition.
The near-field unit is installed on an rotatable frame for
performing THz measurements on the one side and the distance control on the other side.

An accurate distance control is necessary because of the strong dependence of the signal on the
distance between pinhole and sample. It is performed by a
image-processing confocal microscope [Fig.~\ref{fig:scheme}, (3)],
which was developed by Tiziani et al.\ \cite{Tiziani00}.
Therefore the light of an infrared LED (3a) is focused on
the sample via a rotating microlens array (3d). The reflected
light is measured by a CCD camera (3h). Because of the low depth
of focus, which is produced by the microlens array, the reflection
has a maximum if the sample sits exactly in the focus of the last
lens (3f). By moving pinhole and sample relatively to this lens a
three dimensional quantitativ image originates, from which the distance between pinhole and
sample can be measured with a precision of 100~nm within an adequate
time.

\begin{figure}[h]
\centering\includegraphics[width=87mm,clip]{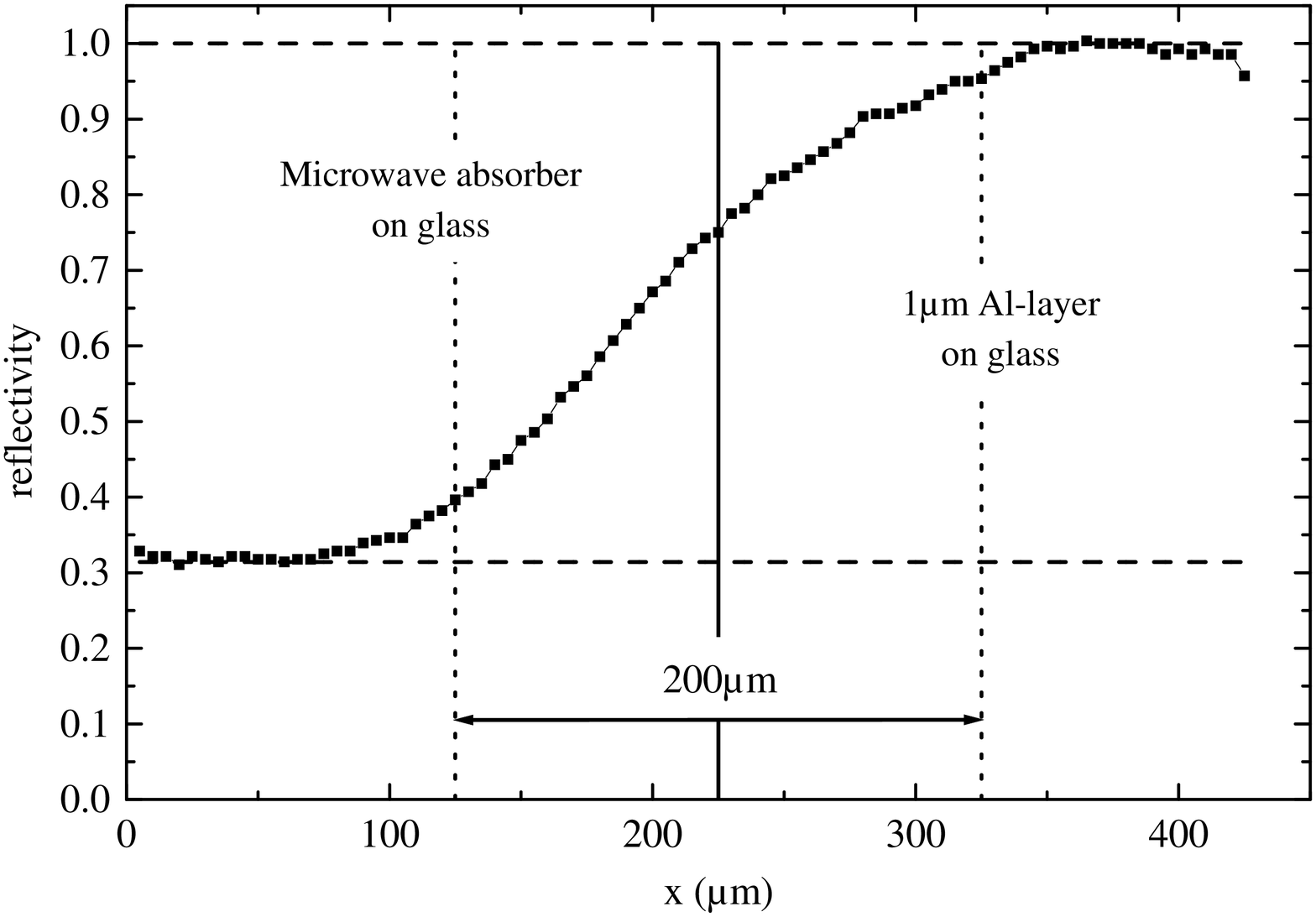}\vspace{3mm}
\caption{\label{fig:resolution}
Scan over an interface between a microwave absorber and silver.}
\end{figure}
Fig.~\ref{fig:resolution} demonstrates the resolution of the
near-field spectrometer. The pinhole was scanned over an interface
between a microwave absorbing film and a silver film, both
deposited on a glass substrate. The increase of reflectivity
within a width of 200~$\mu$m, which corresponds to the diameter of
the pinhole, can clearly be seen. With an used frequency of
180~GHz ($\lambda=1.7$~mm) the near-field ratio $\lambda/d$ is about
10. Additional mathematical treatment (deconvolution) can be used
to improve the spatial resolution considerably. Details of the
experimental setup will be discussed in \cite{Mair02}.

\section{Spectroscopic Measurements}
Due to the strong dependence of the oscillators' output power on
the frequency and due to the frequency-dependent absorption of the
components of the optical spectrometer (lenses, grids, ...),
measurements of absolute value of the reflectivity require  to
perform a reference measurement on an ideal reflecting sample
(e.g. a silver mirror). If only spectroscopic features have to be
identified, difference spectra  are sufficient which were taken of
two materials, at two different sample positions, or two states
different in other respect.

\begin{figure}[b]
\centering\includegraphics[width=70mm,clip]{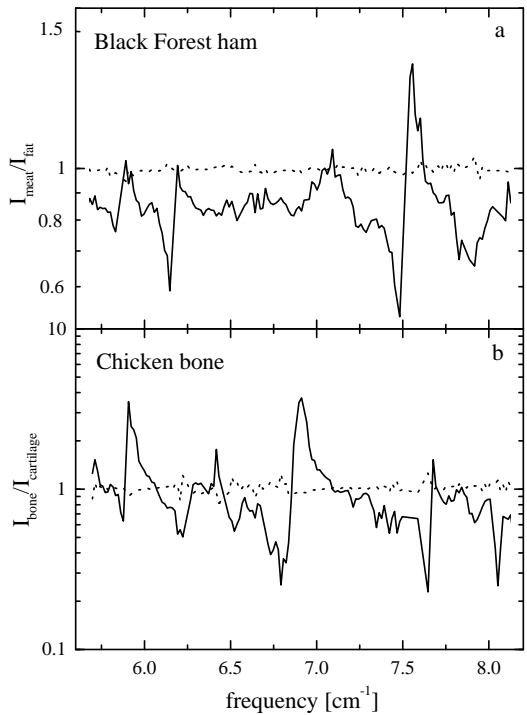}
\caption{\label{fig:spectra} Spectra of two biological samples: a) The solid line shows the reflectivity of the meat part
normalized on the reflectivity of the fat part of Black Forest ham, averaged on three
points each. b) Ratio of the
reflected power of chicken bone and of chicken cartilage.
The dotted lines demonstrates the reproducibility by the 100\% line of the spectrometer.}
\end{figure}

Employing this near-field reflection spectrometer, we have
performed investigations on a variety of biological and medical
samples. As an example Fig.~\ref{fig:spectra}a exhibits the
reflectivity of the meat part of Black Forest Ham relative to the
fat part in the frequency range from 5.5~cm$^{-1}$ to 8.5~cm$^{-1}$. The
circular pinhole used was  200~$\mu$m in diameter in a 10~$\mu$m
thick copper foil. The distance between pinhole and sample was
3.0~$\mu$m. To rule out inhomogeneities of the sample we performed
three measurements on different the meat part as well as on the
fat part and averaged them. The reproducibility of the main
features and the homogeneity of the sample are demonstrated by
dividing data sets taken at different positions, ideally leading
to a 100\% line (dotted line). The spectroscopic distinction
between meat and fat become obvious when the ratio of the spectra
is plotted (solid line).

Similar experiments have been performed on chicken bone which has
a much lower water content and thus a smaller reflection
coefficient. Fig.~\ref{fig:spectra}b shows the ratio of the the
reflected power of the bone and of the cartilage. Again the
spatial resolution is less than 200~$\mu$m.

\section{Spectroscopic Imaging}
In both spectra we find unexpectedly sharp and strong features; in
certain ranges the reflected power is 5 times as high in cartilage
compared to the chicken bone, for instance. This indicates strong
absorption lines which may serve as a spectroscopic signature of
the material. While at presence we cannot reliably assign the
lines, we already want to note the possibility of using these
large differences for imaging. Differential images may be a very
sensitive tool to probe the spatial distribution of certain
molecules with a resolution in the micrometer range. For this
purpose only two distinct frequencies have to be measured which is
easily possible by our spectrometer and which reduces the
measurements time by orders of magnitude. This technique allows to
selectively map out the composition of samples.

\section{Conclusion}
We have developed a new microspectrometer to perform spectroscopic
experiments in the THz range with a large frequency resolution of
10$^7$. The use of  near-field technique enables us to obtain a
spatial resolution down to several micrometers. With a reflection
setup we can investigate biological samples in aqueous solution.
The instrument also provides spectro-images which show the
spatial distribution of biological molecules with a high spatial
resolution. This novel technique has an enormous potential for a
variety of biological and medical applications.

\section*{Acknowledgements}
This study has been carried out with financial support from the Commission of the European
Communities, specific RTD program `Quality of Life and Management of Living Resources',
QLK4-CT-2000-00129  ``Tera-Hertz Radiation In Biological Research, Investigation on Diagnostics and Study on Potential Genotoxic Effects''. It does not necessarily reflects its view and in no way anticipates the Commission's future policy in this area.

\end{multicols}
\end{document}